\documentclass[11pt,twoside]{article}

\begin{document}
\title{\bf A study of the efficiency of the class of $W$-states as a quantum channel}
\author{Satyabrata Adhikari\thanks{Corresponding
Author: Satyabrata Adhikari, e-mail:satyabrata@bose.res.in},
Sunandan Gangopadhyay\\
S.N. Bose National Centre for Basic Sciences, Salt Lake, Kolkata
700 098, India}
\maketitle
\begin{abstract}
\noindent
Recently, a new class of $W$-states has been defined by Agarwal and
Pati \cite{agarwal} and it has been shown
that they can be used as a quantum channel for teleportation and
superdense coding. In this work, we identify those three-qubit
states from the set of the new class of $W$-states which are most
efficient or suitable for quantum teleportation.
We show that with some probability
$|W_{1}\rangle=\frac{1}{2}(|100\rangle+|010\rangle+\sqrt{2}|001\rangle)$
is best suited for teleportation channel in the sense that it does
not depend on the input state.
\end{abstract}
Quantum teleportation \cite{bennett1}  involves the transmission
of an arbitrary qubit from one party to another distant party by
sending two classical bits. However they must initially share one
entangled state.
It can also be combined with other operations to construct advanced
quantum circuits useful for information processing \cite{nielsen}.
The first experimental demonstration of quantum teleportation was
reported by Bouwmeester et.al. \cite{bouwmeester}. Quantum
teleportation is also possible in systems corresponding to
infinite dimensional Hilbert spaces
\cite{vaidman,braunstein,adesso,anno,loock,adhikari1,adhikari2}.\\
The teleportation of quantum entanglement is also known as
entanglement swapping \cite{pan,tan,nielsen,braunstein1}. It
enables two parties (having no common past) to share quantum
entanglement. This protocol may be useful in making nonlocal
correlations over long distances.
Therefore it can play a significant role in quantum communication.\\
Quantum entanglement \cite{einstein}, the heart of quantum
information theory, plays a crucial role in computational and
communicational purposes. Therefore as a valuable resource in
quantum information processing, quantum entanglement has been
widely used in quantum cryptography \cite{brassard,shor}, quantum
superdense coding \cite{wiesner}, quantum teleportation
\cite{bennett1}, remote state preparation of special qubits and
arbitrary quantum states \cite{pati1,bennett2}, and teleportation
of an unitary operator \cite{bruss}. Along with
Einstein-Podolsky-Rosen (EPR) state and
Greenberger-Horne-Zeilinger (GHZ) state, there exists other
entangled states such as $W$-class states and zero sum amplitude
(ZSA) states \cite{pati2} which have substantial importance in
quantum information theory. Tripartite ZSA state may be related to
$W$-class state under stochastic LOCC \cite{dur}.\\
In \cite{karlsson}, it was shown that an unknown qubit can be
teleported using a three particle GHZ state. Joo et.al. \cite{joo}
proposed quantum key distribution (QKD) and partial quantum secret
sharing protocols based on $W$-states of three qubits. Further,
V.N. Gorbachev et.al. \cite{gorbachev1} showed that in contrast to
$|W\rangle=\frac{1}{\sqrt{3}}(|100\rangle+|010\rangle+|001\rangle)$,
$|\tilde{W}\rangle=\frac{1}{2}(\sqrt{2}|100\rangle+|010\rangle+|001)$
can be used as a quantum channel for teleportation of an entangled
state and superdense coding. Recently, Agarwal and Pati
\cite{agarwal} defined a large class of $W$-states that can also
be used as an entanglement resource for teleportation protocol and
superdense coding. Therefore, keeping in mind the importance of
$W$-states, there have been various proposals to prepare these
states in the literature \cite{guo,gorbachev2,biswas}.\\
In this work, we estimate the efficiency of the quantum channel
$|W_{n}\rangle$ defined in \cite{agarwal}. To do this we proceed
in the following way. First we consider a two-qubit entangled state (pure
or mixed) in Alice's place. Alice then send a qubit from her
entangled state to Bob using the special class of three-qubit
quantum channel ($|W_{n}\rangle$) shared by them.
After the completion of the protocol,
if it turns out that the amount of entanglement in Alice's two-qubit state
is retained in the final two-qubit state shared by both Alice and Bob,
then we can say that the efficiency of the teleportation channel
is unity. If the amount of entanglement decreases after completing
the teleportation protocol then the efficiency of the channel is
less than unity. To quantify the above statement, we compute
the concurrence \cite{wootters} which gives a measure
of entanglement. Hence, as it goes
towards zero, one can say that the efficiency of the
channel decreases.

\noindent Let us consider an arbitrary two qubit state
\begin{eqnarray}
|\psi\rangle_{12}=\alpha|00\rangle+\beta|11\rangle\quad,
\quad\alpha^{2}+\beta^{2}=1~.
\label{1}
\end{eqnarray}
For simplicity we assume that $\alpha$ and $\beta$ are real.
As we have mentioned earlier, the above two qubit entangled state
is in Alice's place, i.e. Alice holds
both the particles.\\
The special class of $W$-state is given by
\cite{agarwal}\footnote{Originally there are two
phase factors in $|W_{n}\rangle$
but for simplicity we have ignored them here.}
\begin{eqnarray}
|W_{n}\rangle_{345}=f(n)(|100\rangle+\sqrt{n}
|010\rangle+\sqrt{n+1}|001\rangle)
\label{2}
\end{eqnarray}
where, $n$ is a positive real number and $f(n)$
is the normalization constant
given by
\begin{eqnarray}
f(n)=\frac{1}{\sqrt{(2+2n)}}~.
\label{norm}
\end{eqnarray}
The above three-qubit entangled state serve as a quantum channel
and it is also shared by Alice and Bob. Let us assume that
particle-3 is in Alice's possession and the remaining two
particles (i.e. particles 4 and 5) are with Bob. Our task is to
see the efficiency of this quantum channel in the teleportation protocol.\\
The combined system of five particles can be written as a tensor
product of $|\psi\rangle_{12}$ and $|W_{n}\rangle_{345}$ :
\begin{eqnarray}
|\chi\rangle_{12345}&&=|\psi\rangle_{12}\otimes |W_{n}\rangle_{345}
\nonumber\\
&&=\frac{f(n)}{\sqrt{2}}\left[|\Phi^{+}\rangle_{23}(\sqrt{n}
\alpha|010\rangle_{145}+\sqrt{n+1}\alpha|001\rangle_{145}+\beta|100
\rangle_{145})\right.\nonumber\\
&&\left.+|\Phi^{-}\rangle_{23}(\sqrt{n}\alpha|010\rangle_{145}+\sqrt{n+1}
\alpha|001\rangle_{145}-\beta|100\rangle_{145})
\right.\nonumber\\
&&\left.+|\Psi^{+}\rangle_{23}(\alpha|000\rangle_{145}+\sqrt{n}
\beta|110\rangle_{145}+\sqrt{n+1}\beta|101\rangle_{145})
\right.\nonumber\\
&&\left.+|\Psi^{-}\rangle_{23}(\alpha|000\rangle_{145}-\sqrt{n}
\beta|110\rangle_{145}-\sqrt{n+1}\beta|101\rangle_{145}\right]~.
\label{4}
\end{eqnarray}
Thereafter, Alice performs the measurement on the qubits 2 and 3
in the Bell-basis
$\{|\Phi^{+}\rangle,|\Phi^{-}\rangle,|\Psi^{+}\rangle,|\Psi^{-}\rangle\}$.
If the measurement outcome is $|\Phi^{\pm}\rangle_{23}$, then it indicates
that the five-qubit state $|\chi\rangle_{12345}$ collapsed to the
three-qubit state
$(\sqrt{n}\alpha~|010\rangle_{145}+\sqrt{n+1}
\alpha~|001\rangle_{145}\pm\beta~|100\rangle_{145})$.
Alice then sends the measurement result to her partner Bob by spending
two classical bits. After receiving the classical message from
Alice, Bob performs a von-Neumann measurement on the qubit-5 in
the computational basis $\{|0\rangle,|1\rangle\}$. As a result of
this one-qubit measurement, the three-qubit state reduces to
two-qubit state. Without any loss of generality, let us assume
that von-Neumann measurement outcome is $|0\rangle_5$. Then the
reduced two-qubit state is of the following form
\begin{eqnarray}
|\xi\rangle_{14}= N(\sqrt{n}\alpha|01\rangle \pm \beta|10\rangle)
\label{5}
\end{eqnarray}
where, $N=\frac{1}{\sqrt{n\alpha^{2}+\beta^{2}}}$.\\
Now our task is to calculate the entanglement between the
particles 1 and 4. This calculation of the entanglement gives us
insight about the efficiency of the quantum channel
$|W_{n}\rangle$ in the teleportation protocol.

\noindent To do this, we recall that the concurrence of a pure
state of two qubits is defined as \cite{wootters}
\begin{eqnarray}
C(|\eta\rangle)&=&|\langle\eta|\tilde{\eta}\rangle|
\label{conc}
\end{eqnarray}
where, $|\tilde{\eta}\rangle$ denotes the spin flip of two-qubit pure
state and is defined by
\begin{eqnarray}
|\tilde{\eta}\rangle&=&(\sigma_{y}\otimes\sigma_{y})|\eta^{\ast}\rangle
\label{7}
\end{eqnarray}
with $|\eta^{\ast}\rangle$ being the complex conjugate of
$|\eta\rangle$ when it is expressed in the computational basis
$\{|0\rangle,|1\rangle\}$ and
\begin{eqnarray}
\sigma_{y}\otimes\sigma_{y}=
|11\rangle\langle00|-|01\rangle
\langle10|-|10\rangle\langle01|+|00\rangle\langle11|~.
\label{8}
\end{eqnarray}
With the above definitions at our disposal, we compute
the concurrence for the
state $|\xi\rangle_{14}$ (\ref{5}) and get
\begin{eqnarray}
C(|\xi\rangle_{14})=
\frac{2\alpha\sqrt{n(1-\alpha^{2})}}{(n-1)\alpha^{2}+1}=
C(|\psi\rangle_{12})\left[\frac{\sqrt{n}}{(n-1)\alpha^{2}+1}\right]
\label{9}
\end{eqnarray}
where, $C(|\psi\rangle_{12})= 2\alpha\sqrt{1-\alpha^{2}}$.\\
Note that the concurrence $C(|\xi\rangle_{14})$ depends on the
input entangled state parameter $\alpha$ and $n$.
For any fixed real positive number $n$,
it can be easily observed that as
$\alpha^{2}$ varies from 0 to 1, the concurrence first increases
(from $C(|\xi\rangle_{14})=0$),
attains the maximum value $(C(|\xi\rangle_{14})=1)$ and again
decreases to zero.

\noindent Next we make the
following interesting observations for the case when Alice gets the
Bell-state measurement result $|\Phi^{\pm}\rangle_{23}$ and Bob gets
the von-Neumann measurement result $|0\rangle$ :

\noindent 1. For $n=1$, the final concurrence $C(|\xi\rangle_{14})$
(i.e. concurrence after completing the protocol) is exactly equal to
the initial concurrence $C(|\psi\rangle_{12})$. This means
$|W_{1}\rangle$ serve as the best teleportation channel for any
arbitrary input state (\ref{1}). For $n\neq1$, one can once again have
$C(|\xi\rangle_{14})=C(|\psi\rangle_{12})$ for
some specific states with parameter $\alpha$ taking the
form $\alpha^{2}=\frac{\sqrt{n}-1}{n-1}$~.

\noindent2. There also exist other three-qubit quantum
channels $|W_{n}\rangle$ (which can be used for teleportation) for which
$C(|\xi\rangle_{14})< C(|\psi\rangle_{12})$ and
hence are not as much efficient as the state independent quantum
channel $|W_{1}\rangle$ or the state dependent quantum channels
$|W_{n}\rangle$ ($n\neq1$) with
$\alpha^{2}=\frac{\sqrt{n}-1}{n-1}$. We now classify those states
as follows :

\noindent (i) For $0<n<1$, the range for $\alpha^{2}$ is given by
$0<\alpha^{2}<\frac{\sqrt{n}-1}{n-1}$.\\
(ii)For $n>1$, the range for $\alpha^{2}$ is given by
$\frac{\sqrt{n}-1}{n-1}<\alpha^{2}<1$.\\


\noindent Apart from the above observations, we also note that
whatever be the Bell-state measurement result (obtained by Alice
on qubits 2 and 3), the final concurrence (i.e. the entanglement
of the final state) vanishes if Bob gets the von-Neumann measurement result
$|1\rangle_{5}$. This indicates that the special class
of $W$-states does not act as a good teleportation channel at all.
Furthermore, the concurrence of the final state (when
the Bell-state measurement result of Alice is
$|\Psi^{\pm}\rangle_{23}$ followed by Bob's von-Neumann measurement result
$|0\rangle_{5}$) is given by :
\begin{eqnarray}
C(|\xi\rangle_{14})=
\frac{2\beta\sqrt{n(1-\beta^{2})}}{(n-1)\beta^{2}+1}=
C(|\psi\rangle_{12})\left[\frac{\sqrt{n}}{(n-1)\beta^{2}+1}\right].
\label{9a}
\end{eqnarray}
\noindent Next we extend our idea to the
case of mixed state. We want to
investigate the performance of $|W_{n}\rangle$ as a
teleportation channel when teleporting a single-qubit mixed state.
In particular, we start with the family of Werner state which is of the
form
\begin{eqnarray}
\rho_{12} = p~|\Phi^{+}\rangle\langle\Phi^{+}|+\frac{1-p}{4} I
{}\nonumber\\
=\left(\begin{matrix}{\frac{(1+p)}{4} & 0 & 0 & \frac{p}{2} \cr 0
& \frac{(1-p)}{4} & 0 & 0 \cr 0 & 0 & \frac{(1-p)}{4} & 0 \cr
\frac{p}{2} & 0 & 0 & \frac{(1+p)}{4}}\end{matrix}\right).
\label{werner}
\end{eqnarray}
Note that the concurrence of the above state is $C(\rho_{12})=(3p-1)/2$.
Now considering the above state (\ref{werner}) as an input state and then
teleporting a qubit from it using the same procedure discussed for
pure state, we measure the efficiency of the
quantum channel $|W_{n}\rangle$ by computing the concurrence.
Based on the two-particle followed by one-particle
measurement results, we discuss the following two cases below:\\
Case I: If the Bell-state measurement outcome is
$|\Phi^{\pm}\rangle_{23}$ and the von-Neumann
measurement result is $|0\rangle_{5}$, then after completion of the
teleportation protocol, the two-qubit state will read
\begin{eqnarray}
\rho_{14} = \left(\begin{matrix}{\frac{(1-p)}{8} & 0 & 0 & 0 \cr 0
& \frac{n(1+p)}{8} & \frac{\pm \sqrt{n}p}{4} & 0 \cr 0 & \frac{\pm
\sqrt{n}p}{4} & \frac{ n(1+p)}{8} & 0 \cr 0 & 0 & 0 &
\frac{n(1-p)}{8}}\end{matrix}\right). \label{12}
\end{eqnarray}
The concurrence for the above state described
by the density matrix $\rho_{14}$ is therefore given by
\begin{eqnarray}
C(\rho_{14})&=&\left\{
\begin{array}{cl}
\frac{4\sqrt{n}(3p-1)}{(n+1)^{2}}=
C(\rho_{12})\frac{8\sqrt{n}}{(n+1)^{2}}~,~~~~if~~p>\frac{1}{3}\cr
=0~,\qquad~~~~~~~~~~~~~~~~~~~~~~~~~if~~p\leq\frac{1}{3}.\cr
\end{array}
\right.
\label{13}
\end{eqnarray}
Now since $C(\rho_{14})\leq C(\rho_{12})$, hence we have the following
inequality satisfied by $n$
\begin{eqnarray}
n^{4}+4n^{3}+6n^{2}-60n+1\geq 0.
\label{ineq}
\end{eqnarray}
The above relation explicitly gives us information about those
three-qubit quantum channels $|W_n\rangle$
which can be used as a teleportation
channel for family of Werner states with parameter $p>\frac{1}{3}$. If
the inequality (\ref{ineq}) becomes an equality then it provides us
information about the quantum channel with unit efficiency . If
$p\leq\frac{1}{3}$, then it can be easily checked that
the concurrence vanishes and hence the $|W_n\rangle$
quantum channel fails to teleport.\\

\noindent Case II: If the Bell-state measurement outcome is
$|\Psi^{\pm}\rangle_{23}$ and the von-Neumann
measurement result is $|0\rangle_5$, then after completion of the
teleportation protocol, the two-qubit state will read
\begin{eqnarray}
\rho'_{14} = \left(\begin{matrix}{\frac{(1+p)}{8} & 0 & 0 &
\frac{\pm \sqrt{n}p}{4} \cr 0 & \frac{n(1-p)}{8} & 0 & 0 \cr 0 & 0
& \frac{ (1-p)}{8} & 0 \cr \frac{\pm \sqrt{n}p}{4} & 0 & 0 &
\frac{n(1+p)}{8}}\end{matrix}\right).
 \label{15}
\end{eqnarray}
In this case, the concurrence is exactly equal to the concurrence
given in equation (\ref{13}) and hence the fact about the quantum
channel $|W_n\rangle$ remains the same as in Case I.\\

\noindent But unlike the above two cases, we observe that for any
Bell-state measurement made by Alice on qubits 2 and 3, if the
von-Neumann measurement result (made by Bob) is $|1\rangle_5$,
then the concurrence of the resultant state vanishes and hence
once again the $|W_n\rangle$-state fails to act as a teleportation
channel.\\
To summarize, we have presented a protocol that measures the
efficiency of the class of $W$-state defined in \cite{agarwal}. We
observe that there exists two
types of $|W_n\rangle$-state which can serve as a
quantum channel for entanglement swapping. They can be classified
as: (i) State independent channel and (ii) State dependent
channel. $|W_{1}\rangle$ is regarded as a state independent
channel because with some probability it helps in swapping the
initial entanglement to the final two-qubit state
for any arbitrary two-qubit input states. On the other hand, there
exist quantum channels which retains the initial entanglement in
the final two-qubit state only for few input
two-qubit states. Hence we call these type of channels as state
dependent channel. We finally extend our analysis to the mixed state also.


\begin{thebibliography}{99}
\bibitem{bennett1}  C. H. Bennett {\it et al},
Phys.Rev.Lett. \textbf{70}, 1895 (1993).\
\bibitem{nielsen} M. A. Nielsen and I. L. Chuang,
\textit{Quantum Computation
and Quantum Information}, (Cambridge University Press, Cambridge,
2000).\
\bibitem{bouwmeester} D. Bouwmeester {\it et al}, Nature \textbf{390}, 575
(1997).\
\bibitem{braunstein1} S. L. Braunstein and A. K. Pati,
\textit{Quantum Computation with continuous variables}, Kluwer
Academic Publishers, Dordrecht, 2003).\
\bibitem{vaidman} L. Vaidman, Phys. Rev. A \textbf{49}, 1473 (1994).\
\bibitem{braunstein} S. L. Braunstein and H. J. Kimble,
Phys. Rev. Lett. \textbf{80}, 869 (1998).\
\bibitem{adesso} G. Adesso and F. Illuminati,
Phys. Rev. Lett. \textbf{95}, 150503 (2005).\
\bibitem{anno} F. Dell'Anno {\it et al},
Phys. Rev. A \textbf{76}, 022301
(2007).\
\bibitem{loock} P. van Loock and S.L.Braunstein,
Phys. Rev. Lett. \textbf{87}, 247901
(2001).\
\bibitem{adhikari1} S. Adhikari {\it et al},
Phys. Rev. A \textbf{77}, 012337 (2008).\
\bibitem{adhikari2} S. Adhikari, e-print quant-ph/0802.2156.\
\bibitem{einstein} A. Einstein {\it et al},
Phys. Rev. \textbf{47}, 777
(1935).\
\bibitem{brassard} C. H. Bennett and G. Brassard,
\textit{Proceedings of the IEEE
international Conference on Computers, System, and Signal
processing}, Bangalore, India, (IEEE, New York, 1984), pp.
175-179.\
\bibitem{shor} P. W. Shor and J. Preskill,
Phys. Rev. Lett. \textbf{85},
441 (2000).\
\bibitem{wiesner} C. H. Bennett and S. Wiesner,
Phys. Rev. Lett. \textbf{69}, 2881
(1992).\
\bibitem{pati1} A. K. Pati, Phys. Rev. A \textbf{63}, 014320-1
(2001).\
\bibitem{bennett2} C. H. Bennett {\it et al},
Phys. Rev. Lett. \textbf{87}, 077902 (2001).\
\bibitem{bruss} S. F. Huelga {\it et al},
Phys. Rev. A \textbf{63}, 042303 (2001).\
\bibitem{pati2} A. K. Pati, Pramana J. Phys. \textbf{59}, 217 (2002).\
\bibitem{dur} W. Dur {\it et al},
Phys. Rev. A \textbf{62}, 062314 (2000).\
\bibitem{joo} J. Joo {\it et al}, e-print quant-ph/0204003.\
\bibitem{gorbachev1} V. N. Gorbachev {\it et al},
Phys. Lett. A \textbf{314}, 267 (2003).\
\bibitem{agarwal} P. Agarwal, and A. K. Pati,
Phys. Rev. A \textbf{74}, 062320 (2006).\
\bibitem{guo} G. C. Guo, and Y. -S. Zhang,
Phys. Rev. A \textbf{65}, 054302 (2002).\
\bibitem{gorbachev2}  V. N. Gorbachev {\it et al},
Phys. Lett. A \textbf{310}, 339 (2003).\
\bibitem{biswas}  A. Biswas, and G. S. Agarwal,
J. Mod. Opt. \textbf{51}, 1627 (2004).\
\bibitem{wootters}  W. K. Wootters,
Phys. Rev. Lett. \textbf{80}, 2245 (1998).\
\bibitem{karlsson} A. Karlsson, and M. Bourennane,
Phys. Rev. A \textbf{58}, 4394 (1998).\
\bibitem{pan} J. -W. Pan {\it et al},
Phys. Rev. Lett. \textbf{80}, 3891 (1998).\
\bibitem{tan} S. M. Tan,
Phys. Rev. A \textbf{60}, 2752 (1999).\
\end{thebibliography}
\end{document}